# RSCM Technology for Developing Runtime-Reconfigurable Telecommunication Applications

Kamaledin Ghiasi-Shirazi[*], Mahdi Mohseni, Majid Darvishan, and Reza Yousefzadeh[§]

*Abstract*—Runtime reconfiguration is a fundamental requirement of many telecommunication applications which also has been addressed by management standards like CMIP, 3GPP TS 32.602, and NETCONF. Two basic commands considered by these standards are CREATE and DELETE which operate on managed objects inside an application. The available configuration management technologies, like JMX, OSGi, and Fractal, do not support the CREATE and DELETE reconfiguration commands of the telecommunication standards. In this paper, we introduce a novel technology, called RSCM, for development of runtime reconfigurable applications complying with the telecommunication standards. The RSCM subagent takes the responsibility of loading the application from the configuration file, executing the runtime reconfiguration commands (including CREATE and DELETE), enforcing validity of the configuration state, and updating the configuration file according to the latest reconfiguration commands. We exploit the modular and object oriented features of the XML technology for storing the configuration state of a program in a configuration file. The software development process is tailored such that the design of XML schemas of managed classes is performed parallel to the design of software classes. In addition, a novel programming approach based on indirect referencing is proposed which allows safe and almost immediate deletion of managed objects at runtime. This indirect referencing mechanism affects the implementation of associations in class diagrams and prevents methods of a class to use the "this" pointer freely. The RSCM technology has been successfully used in several commercial telecommunication applications; including an SMS service center, an SMS gateway, and an SMS hub.

I. INTRODUCTION

MANY telecommunication applications, e.g. routers, gateways, and hubs, provide such critical services where it is completely unacceptable to stop them for reconfiguration; so they demand runtime reconfiguration. Several management standards, e.g. CMIP[1], 3GPP TS 32.602[2] ,and NETCONF [3], have been proposed for the runtime reconfiguration of the network and telecommunication applications. These standards assume that an application consists of a set of managed objects which are organized as a tree called MIB (Management Information Base). The common runtime reconfiguration commands specified by these standards are: GET, SET, CREATE, and DELETE. The GET command gets the values of the configuration parameters, the SET command updates the configuration parameters, the CREATE command creates a new managed object, and the DELETE command deletes a managed object.

Among these runtime reconfiguration commands, the CREATE and DELETE commands are much more difficult to implement. At execution of the CREATE command, the new managed object must be completely initialized: the configuration variables, references from this object to other objects, and references from other objects to this object, all should be initialized with appropriate values. Since a managed object may contain several other managed objects, the implementation of the CREATE command should also take the responsibility of creating and initializing the subtree of the contained managed objects as well. Furthermore, for the CREATE command to be meaningful, one should specify what types of managed objects can be created and where these managed objects can be added to the MIB tree. The difficulty of implementing the DELETE command lies in the fact that there may be many references to the deleting managed object. On the one hand, deleting an object, while being referenced by other objects, leads to dangling pointers and consequently segmentation fault errors. On the other hand, postponing the deletion of a managed object until the references are released may delay the execution of the DELETE command for an unlimited time.

---

[*] Corresponding author.
[§] K. Ghiasi-Shirazi, is with the Computer Engineering Department, Ferdowsi University of Mashhad, Azadi Sq., Mashhad, Khorasan Razavi, Iran (e-mail: k.ghiasi@um.ac.ir).
M. Mohseni, M. Darvishan, and R. Yousefzadeh are with PeykAsa Corporation, No. 10, 4th Alley, Sadeghi Street, Azadi Avenue, Tehran, Iran (e-mail: [mahdi.mohseni,majid.darvishan,reza.yousefzadeh]@peykasa.ir).

The task of executing the runtime reconfiguration commands includes many application-independent subtasks such as communicating management packets over the network, validating the reconfiguration commands, mapping the name of an object to its reference within the application, calling methods of managed objects (instead of calling a global function), and updating the configuration file according to the latest configuration state. Implementations of telecommunication standards like 3GPP TS 32.602 or NETCONF (e.g. libnetconf[4]) solely invoke a callback function, asking the application developer to perform the CREATE and DELETE commands. Various technologies have been developed that encapsulate some/all of these subtasks in a reusable software library called the subagent. The JMX technology [5], the OSGi configuration admin service[6], and the Fractal framework [7] propose the most important general purpose subagents for configuration management. The JMX and OSGi technologies are confined to the java programming language, but the Fractal framework is available both in java and C programming languages. Menten[8] developed a subagent specifically designed for management of network devices with limited resources that is specially powerful in representing and validating the configuration state using the XML technology.

However, none of these subagents support the CREATE and DELETE reconfiguration commands as specified by the telecommunication standards. While the JMX technology has provisioned mechanisms for setting and getting configuration variables and performing operations on managed objects, it does not provide any mechanism for creating or deleting managed objects at all. The OSGi configuration admin service introduces a special kind of component, called Managed Service Factory, instances of which can be used for creating and deleting managed objects of a specific managed class. However, since the OSGi configuration admin service does not support the tree structure of MIB for managed objects, the supported CREATE and DELETE operations differ considerably from those specified by the telecommunication standards. The Fractal programming framework defines a well-designed set of interfaces for Fractal components, but its subagent does not exploit these interfaces for implementing the CREATE and DELETE commands. Menten's subagent[8] finds the parent object and sends it an XML fragment containing the necessary information for creating the child object. However, it leaves the task of creating the subtree of objects, as specified by the XML fragment, to the parent object. In addition, none of these technologies address the issue of how a managed object can be deleted in a safe and immediate way.

In this paper, we introduce the RSCM (Runtime Software Component Management) technology which performs the complete DELETE and CREATE operations as specified by the standards with a minimum intervention from application programmer. The RSCM subagent exploits some simple programming interfaces, implemented by the managed objects, to fulfill the CREATE and DELETE reconfiguration commands. In addition, the proposed technology guarantees the safe and almost immediate deletion of managed objects. Since the proposed technology is targeted for the telecommunication applications, we assume that, as specified by the telecommunication standards, the relationships between managed objects can be represented as a MIB tree. We store the configuration state of the application in an XML document and use XML Schema to validate the integrity of the configuration state. In addition, the RSCM technology provides programming interfaces which can be used for defining arbitrarily complex validation rules. The RSCM subagent also takes the responsibility of updating the configuration file according to the latest reconfiguration commands.

We also pay special attention to the software engineering issues[9] of developing runtime reconfigurable applications. Specifically, we address the issue of how the proposed technology can be used in developing reusable software libraries with built-in runtime reconfigurability. Another novelty of the proposed technology is making connections between the object oriented design in the programming language and the XML Schema of the configuration file.

We have implemented the RSCM subagent in C++ and java programming languages. A CLI (Command Line Interface) complying with ITU-T Z.315[10] standard and a web-based interface complying with the 3GPP TS 32.603[11] standard have been developed for management of the applications developed by the RSCM technology. Table 1 compares the RSCM technology with JMX, work of Menten[8], Fractal and libnetconf[4].

The rest of the paper proceeds as follows. In section II, we introduce the RSCM technology and explain how it addresses the key issues of developing runtime reconfigurable applications. In Section III, we report some experiments on development of real-world telecommunication applications with the RSCM technology. We conclude the paper in Section IV.

## II. THE PROPOSED TECHNOLOGY

### A. Subagent's Management Operations

After studying various network and telecommunication management standards, we concluded that the subagent should support the following basic management operations:

*1) Operations for investigating the MIB tree*

ListChildren: This operation takes the name of a parent managed object as input and returns the names of its children; if the input is the empty string, then the root objects are returned[1].

*2) Configuration management operations*

ListConfigurationParameters: This command returns the list of the configuration parameters of a managed object.

SetConfigurationParameter: This command changes the values of a selected subset of the configuration parameters of a managed object.

GetConfigurationParameter: This command returns the values of a selected subset of the configuration parameters of a managed object.

GetChildCreationSignature: To CREATE a new managed object A, one needs to know the configuration parameters of A, the managed objects that are contained in A, and recursively the configuration parameters and child managed objects of any contained object. This command gets the name of a parent object and returns the signature of the child managed objects which can be added to it.

CreateManagedObject: This command takes the name of a parent managed object and the configuration parameters of a new child and creates a new child with the specified configuration parameters and adds it to the parent managed object. If the child object itself contains more nested child managed objects, the subagent creates them recursively.

DeleteManagedObject: This command takes the name of a managed object and deletes it from the application.

*3) Action Operations*

ListActions: This command returns the list of the supported actions of a managed object.

GetActionSignature: This command gets the name of a managed object along with a valid action name and returns the signature for calling that action as a list of parameters and types.

DoAction: This command executes an action on a managed object.

Although the main focus of this paper is on the runtime reconfiguration, the proposed technology supports fault and performance management operations as well. The subagent provides fault management programming interfaces for the application developers to send notification of new alarms or clearance of previously reported alarms to the agent. Internally, the subagent uses the SNMP protocol to send these alarms to the agent. The subagent provides the following performance management operations.

*4) Performance Management Operations*

ListMonitoringVariables: This command returns the list of the monitoring variables of a managed object.

GetMonitoringVariables: This command gets a list of monitoring variable names and returns their values.

*B. Constructing Managed Objects According to the Configuration File*

Supporting the CREATE and DELETE reconfiguration commands means that the set of managed objects which are present in a running application can be modified at runtime. Furthermore, the configuration file should be updated according to the latest reconfiguration commands so that the application can resume from the latest configuration state after an unwanted crash. Thus, one of roles of the RSCM subagent is to initially load the application and create managed objects based on the configuration file.

*C. Categorization of Managed Objects into Simple and Tabular*

To be able to CREATE a managed object in a running application we should at least answer the following questions:
1- Which objects can have new children?
2- What types of objects can be added to a parent managed object?
3- What is the relative distinguished name of a newly created managed object? This name would be used by the consequent management commands.

To answer the first question, we categorize the managed objects into two types: simple and tabular. Simple managed objects are those objects which have a fixed set of child managed objects. In contrast, tabular managed objects are those objects which their list of child managed objects can be modified by runtime reconfiguration commands.

To answer the second question, we assume that all children of a tabular managed object are of the same class. In addition to declaring the signature of managed objects that can be added to a tabular managed object, this assumption allows a tabular display for children of tabular managed objects (see Figure 17). Note that since tabular managed classes are added artificially to the design of a program, it is natural to assume that they have no configuration variables. This assumption generates a much cleaner representation of tabular managed objects in the configuration file and user interfaces. It must be mentioned that these assumptions only organize the managed objects and don't impose any real restriction on the manageability of an application. In general, any initial set of managed classes can be converted to a new set that satisfies the above assumptions. Assume that, in an arbitrary initial design, the managed class P is composed of a set of configuration parameters $P_g$, a set of fixed children $C_1,\ldots,C_m$, and an arbitrary number of children of types $S_1,\ldots,S_n$. The modified class P would be a simple managed class with the configuration parameters $P_g$ and fixed children $C_1,\ldots,C_m$ and $T_1,\ldots,T_n$, where, for $1 \leq j \leq n$, $T_j$ is a tabular managed class containing

---

[1] Standards emphasize that the MIB is a tree. However, the containment relationship between managed objects of an application may lead to a forest. So, within an application we accept that the managed objects form a forest with several roots. By considering the application as a managed object which contains all of its managed objects, we arrive at a tree structure which complies with tree structure view of the standards.

arbitrary number of children of type $S_j$ (see Figure 1 ).

Now, we answer the third question. We assume that the child managed objects which can be added to a tabular managed object have a distinguishing key configuration variable. This assumption is taken from the telecommunication naming standards such as ITU-T Recommendation X.500[12] or 3GPP TS 32.300[13]. When a unique key attribute does not come out naturally, an artificial unique attribute, say id, is added to the configuration parameters of the child managed class. Two other possibilities to refer to a new child object are:
1- Numbering the objects according to their order in the configuration file: This approach becomes problematic when multiple managers are simultaneously managing an application. If one manager decides to delete a child and another decides to modify a child which comes after the deleting child, then the execution of the delete operation modifies the number associated to the other child and so the management command of the second manager would be applied to a wrong object. Note that functioning in a transactional way does not solve the problem. The reason is that when issuing a command, the managers are numbering the objects according to their current view which becomes inconsistent after CREATE and DELETE operations. On the other hand, extending the duration of a transaction to the whole time of displaying the MIB tree is equivalent to saying that managers cannot be executed simultaneously.
2- Using compound keys to identify the children: The main problem with this approach is that the naming standards, like ITU-T Recommendation X.500 or 3GPP TS 32.300, don't support compound keys.

*D. Object-oriented and Modular Design*

One important feature of the RSCM technology is that it is meticulously designed to be used in a modular and object oriented software development process[14]. In the RSCM technology, the definition of a managed class in the software and the XML Schema are viewed as two manifestations of the same entity. So, the object oriented design of managed classes in XML Schema parallels their design in the programming language. Of special importance is the inheritance relationship between classes which is also modeled using XSD (XML Schema Definition) language[15, 16]. In addition, the RSCM technology realizes the modularity principle by supporting the development of software libraries with built-in manageability. The management code for the managed classes of each software library goes to the same library. This frees the application developer from the task of implementing the management interfaces of the managed objects of imported software libraries.

*E. Validating Initial Configuration File and Reconfiguration Commands with XML Schema*

To increase the robustness of the application, each reconfiguration command should be validated before being sent to a managed object. To have the maximum power in expressing the validation constraints we allow the developers to express these constraints in code. However, many constraints on the configuration parameters (like the type of the configuration parameter, its minimum value, its length, and the regular expression of valid values) and many constraints between the managed objects (like the uniqueness of the key attribute between children of a tabular managed object) can be defined outside the source code using the XSD language. In addition, there are many graphical tools which make the task of generating XSD files much easier.

One of the most denounced issues with the XML Schema is its low readability [17]. To remedy this issue, NETCONF standard allows the human users to define the constraints using a new data modeling language called Yang which can then be automatically converted to an XML Schema [18]. We believe that although the readability is a severe issue when the XML Schema is defined in a single large file, it is not an issue for small XML Schema documents. In the RSCM technology, each managed class would have a separate XSD document which is usually very small. The complete XSD document associated with the whole XML configuration file is then automatically generated from these tiny XML Schema documents and a sample XML configuration file[2].

*F. Augmenting Class Definition with XML Schema Definition*

Writing an XML Schema that verifies the validity of the configuration file is a huge task which ideally should be distributed among the developers. The managed classes of an application may be developed by the application developers themselves or the developers of the software libraries. In coordination with the principle of modularity, the developers of each managed class should also write the corresponding XML Schema. The rule is that each managed class should have an accompanying XML Schema document in the same module.

*G. Safe and Almost Immediate Deletion of Managed Objects*

A technology for runtime reconfiguration would be incomplete unless it addresses the issue of how the managed objects are to be deleted safely and almost immediately while they are currently being used by the application. A command for deletion of a managed object (and all its descendants) may be issued at any time while there are many references to it.

In the RSCM technology, we have provisioned a mechanism which guarantees the safe and fast deletion of managed objects. The key idea is to use indirect referencing to avoid dangling pointers when an object is deleted. We assign to each managed object, at the time of creation, a unique positive number called ObjectId. The subagent stores a mapping from ObjectIds to

---

[2] The reader may wonder why a sample XML document is used for generating the unified XSD document. The answer is that in the process of generating the unified XSD document, we replace the base classes specified in the XML Schema documents with the associated derived classes specified in the sample XML document. This allows us to have the flexibility of using inheritance in defining classes in XML Schema in development phase while at the same time imposing strict type checking of the derived classes in the final XSD document of the application at deployment phase.

references to managed objects. This mapping can return the reference associated to an ObjectId in *O(1)* time. To delete a managed object, the RSCM subagent first removes the mapping between the ObjectId and the managed object and associates the ObjectId to the null pointer (see Figure 2). Then it decreases the reference count of the deleting managed object which would lead to the actual deletion after all direct references to it have been released. Therefore, when requesting a reference to an object from its ObjectId, the programmers should consider the case that the managed object has been deleted and a null pointer is returned. The appropriate action that ought to be taken after receiving a null pointer completely depends on the nature of the deleted object. Here we give two exemplary situations from the WordReplacer sample application of Section III.C in which the WordReplacementRule is a managed class that represents rules for replacing some word by another and the SmppConnection is a managed class that represents an SMPP protocol network connection. The first situation is when a rule for substituting word A with B is deleted. In this case the application can simply leave instances of word A unaffected. Here, the deletion of the managed object (an instance of WordReplacementRule) can simply be regarded as if it never existed and the application continues its normal work. The second situation is when an instance of SmppConnection is deleted. The SmppConnection contains several threads which are responsible for receiving and sending messages. When an instance of SmppConnection is deleted, the threads within it will receive a null pointer after requesting a reference to this SmppConnection object. Here, the returned null pointer means that the SmppConnection has been deleted and the threads are terminated. In some situations the requirement is that the managed objects that are referring to a deleted managed object (and cannot continue their work without it) should be kept in memory until one of the following cases happens

- Either the deleted managed object is created again in which case the application can continue its normal work.
- Or a time out event is triggered in which case the referring managed objects are deleted as well.

Since the ObjectId of the deleted instance and the newly created one differ, this situation requires that we refer to managed objects by their unique identifier within the parent tabular managed object. This allows the application to continue its work with the deleted managed object when a new instance of it is created (i.e. in a DELETE then CREATE scenario).

Although indirect referencing provides a safe way for storing a reference to a managed object, a direct reference is required for accessing its attributes and methods. The RSCM subagent provides a method that returns a smart pointer to an object given its ObjectId. However, if this direct reference is not released in a short time, then the reference counting method inside smart pointer would postpone the actual deletion for an unacceptable long time. Therefore, to ensure that a managed object would be deleted in a reasonable time, all direct references to managed objects should be released in a short time. The reasonable time is something that must be determined by the application and it may even depend on the ManagedClass itself. For example, it may be acceptable for objects of some type to be deleted 5 minutes after the delete command is issued. In that case the programmers should release and regain smart pointers to these objects at most at 5 minute intervals. If objects of another type should be released at most one second after issuing the delete command, then the programmers should release/regain smart pointers to these objects in at most one second intervals. Usually, there is no hard limit for the execution of a command and so it is not a real issue to define the duration of "short time" exactly. The mechanism is similar to the multitasking in Windows 3.1 in which it was possible to run several tasks provided that they returned the control to the operating system after a reasonably short time. It must be mentioned that, in practice, the requirement of almost immediate deletion of managed objects is not so rigid and deletion of a managed object in say 5 seconds is completely acceptable for the support team. When using the RSCM technology, the following rules should be obeyed by programmers:

1- All association (and consequently aggregation and composition) relationships between managed classes should use the ObjectId as the reference. Use of direct referencing for representing associations between managed objects should be avoided altogether. The preferred way for referring to children of a tabular managed object is to store their unique identifier. This guarantees that the reference would be valid even if the managed object is removed and added again.
2- Direct references to managed objects, which are obtained from ObjectIds, should only be used by a single thread. If a managed object should be accessed by multiple threads, each thread should obtain its own local reference from ObjectId. This becomes very important when a managed object is used implicitly. For example, if a managed object creates a thread, it is customary that the programmers use the "this" pointer freely. However, the true approach is to obtain a reference to "this" from its ObjectId.
3- References to managed objects (even "this") should not be used for a long period of time. If a reference to a managed object is needed for a long period of time, the task ought to be broken into short periods of time where at each time period the reference is released and regained using ObjectId of the managed object. For example, if the object is being used inside an infinite loop, where each iteration lasts less than a second, then the reference to the managed object should be obtained from ObjectId and released at each iteration. In fact, since references are smart pointers, they will be released automatically at the end of the block of loop. In the case that each iteration of the loop lasts longer than one second, the smart pointer ought to be released and regained several times during each iteration of the loop. Note that the exact definition of short/long periods of time are application-specific and even may differ between managed classes.

We now show that the above-mentioned mechanism results in safe and almost immediate deletion of managed objects. Assume that a DELETE command for object A with ObjectId=15 is issued. Let, at the time of issuing the DELETE command, $O_1, \ldots O_n$ be all objects that have an indirect reference to A by its ObjectId=15 and let $t_1, \ldots, t_m$ be threads that have direct references to it. After issuing the DELETE command, the RSCM subagent associated ObjectId=15 with the null pointer and calls the release method on object A. So, any subsequent request to get a reference to object A will be answered by a null pointer. By assumption, all threads $t_1, ..., t_m$ will release their references in a short period of time. Since all references to object A would be released in a short time and no further references would be created, the object A will be actually deleted almost immediately. In addition, the reference counting mechanism ensures that the object would not be deleted until all references to it are released. Note that, at this point, objects $O_1, ... O_n$ are still indirectly referring to the deleted object A with ObjectId=15. However, this does not lead to a crash of the application, since any request to get a reference to an object with ObjectId=15 will be answered by null and so the application would be notified that the object has been deleted. In practice, this mechanism was mostly used when deleting a managed object which is composed of several other managed objects. In this case, receiving a null pointer means that the parent of a subtree of managed objects has been deleted and all the subtree should be deleted as well.

## III. EXPERIMENTS

We implemented the RSCM subagent in C++ and java programming languages. We used the Apache Xerces library, which is available both in C++ and java, to process the XML files and to validate the XML documents against XSD files. The experiments reported in this section are conducted to evaluate RSCM in achieving the following goals
1. Usability for the development of runtime-reconfigurable telecommunication applications
2. Usability in object oriented software development
3. Usability in developing runtime-reconfigurable software libraries
4. Successful execution of the CREATE command
5. Safe and almost immediate execution of the DELETE command

Perhaps the best evidence that supports the success of RSCM in achieving all of these goals is its successful application in developing industrial runtime-reconfigurable telecommunication applications. In fact, the RSCM technology has been successfully used in development of more than 10 telecommunication applications, including an SMS Service Center, an SMS Gateway, and an SMS Hub. These applications have been deployed in several short message sites and are being used for more than 4 years. The test team of the Peykasa Company performs different tests to ensure that applications have the required quality to be deployed in telecommunication sites. Due to the importance of the runtime reconfigurability of the applications and the financial penalties that are enforced by telecommunication companies in the case an application is stopped, automatic tests are performed on the applications to ensure that they are completely reconfigurable at runtime. Since these industrial applications have many irrelevant and complicated issues, it is almost impossible to give a detailed report on how these real-world applications have been developed using the RSCM technology. In Section III.A we report some statistics of the usage of this technology in a commercial SMS Service Center.

In order to give the reader a deeper understanding of how runtime-reconfigurable software libraries and standalone applications are developed by RSCM and to give a more detailed evaluation of RSCM in achieving the above-mentioned goals, we report in-depth experiences on developing two simplified software components (a software library and an executable application) by the RSCM technology. As the first example, in Section III.B we introduce a manageable server-side networking library for SMPP protocol with the ability to manage the connection objects. As the second example, in Section III.C we use this SMPP networking library to develop a WordReplacer application which receives messages from an SMPP client and replaces the words in the messages according to some replacement rules. These experiments confirm the usability of RSCM technology in modular and object oriented software development (Goals 2 and 3). We will show that the WordReplacer application would not be involved in any configuration management operation that belongs to SMPP protocol and networking. In addition to prove the success of RSCM in achieving goals 1, 2 and 3, the SMPP networking example shows one of our toughest challenges in applying RSCM to the management of telecommunication applications. In telecommunication applications, management of network connections is of particular importance. However, in ordinary design of networking libraries, the sever-side connections are objects that are created after a client is connected and destroyed after the client disconnects. We explain how we designed the SMPP networking library such that the connections became manageable. Finally, in Sections III.D and III.E we show how the WordReplacer application can be managed with our CLI and web-based managers. We show examples of the execution of CREATE and DELETE commands on the WordReplacer application. Specifically, we conduct and experiment in which a highly active managed object will be deleted safely and almost immediately (Goals 4 and 5). The implementation details of these simplified examples along with images and videos of management of the WordReplacer application using our CLI and web-based management systems are available at the first author's homepage[3].

### A. Using RSCM for management of an SMS Service Center

A Service Center (SC)[19] is a telecommunication application which is responsible for store-and-forwarding of short messages. PeykAsa Company has developed an SC that supports delivering of 10000 short messages per second. Since restarting this application would lead to loss of hundreds of thousands of short messages, the application should be reconfigured at runtime.

---

[3] http://profsite.um.ac.ir/~ghiasi/publications/RSCM/index.html

The PeykAsa SC is developed in C++ programming language using the RSCM technology. The application uses 10 other libraries that are also developed by the RSCM technology. This application contains 13 tabular managed objects and 60 simple managed objects which contain 450 configuration variables plus 474 monitoring variables. The most challenging CREATE and DELETE operations are due to a simple managed object consisting of 4 sub-managed objects and a total of 79 configuration variables. This is a highly dynamic managed object that is associated to a network connection and creates several threads for processing the packets of this connection. Due to the importance of the runtime reconfigurability of this application and the financial penalties that are enforced by telecommunication companies in the case of a restart, numerous tests are automatically performed to ensure that the application is completely reconfigurable at runtime. After four years of deployment in several telecommunication sites, it can be said that the use of the RSCM technology for management of this SC was completely successful and the application is being managed using our CLI and web-based management tools.

*B. A Networking Library with Manageable Connections for SMPP Protocol*

In this example, we explain the process of producing a server-side SMPP networking library called SmppNetworking. Here, one important design goal is to have manageable networking connections. As mentioned earlier, the RSCM technology can only manage objects which are created at initial load of the application from the configuration file or by a CREATE reconfiguration command. However, a connection object is usually created when a client connects and is deleted after the network connection is closed. Thus, in the usual design the connection objects are created by an event and so cannot be managed by the RSCM technology. Solving this problem in the client side is easy: it is enough to permanently associate a connection object with each client-side network connection which continues to exist even when the network connection is closed. However, management of server-side connections is more subtle. For those network protocols where the clients cannot be identified after losing the network connection, like HTTP, the lifetime of a server-side connection object is restricted to the lifetime of the associated network connection. However, in the SMPP protocol clients send a bind packet immediately after the network connection is established. The bind packet contains a SystemId field which serves as the identity of the client. In our modified design, the SMPP server creates a connection object for each valid client that is listed in the configuration file (or created using CREATE command) and uses these objects when the associated clients actually connect. This way, the lifetime of the connection objects would be independent of whether the network connection is established or not.

The detailed design of the SmppNetworking library is as follows. For each valid client, we consider an instance of the SmppExternalClient class which contains the client's information like SystemId and Password. The SmppExternalClient class represents an ESME (External Short Messaging Entity) in the SMPP protocol. Each object of type SmppExternalClient contains a child of type SmppConnection whose lifetime is the same as its parent. The creation/deletion of SmppConnection objects, like any other managed object, is performed by the reconfiguration commands and the establishment of the actual network connection has no effect on it. After a client connects, we create a temporary non-manageable object of type SmppConnection. The configuration parameters of this temporary connection are cloned from a manageable SmppConnection called the defaultConnectionTemplate. According to the SMPP protocol, when a client connects to a server it should send a Bind packet which contains a SystemId field that identifies the associated SmppExternalClient object. After receiving the Bind packet, we copy the socket information of the temporary SmppConnection to the actual manageable SmppConnection that is associated with the client and delete the temporary SmppConnection (see Figure 3).

Main managed classes of the SmppNetworking library are:
1- SmppServer: The SmppServer is the root object which encompasses all other objects associate with a SMPP server. SystemId of the server and the SessionInitTimeout (which according to the SMPP protocol is the longest time where a client is allowed to delay between establishment of the connection and sending the Bind packet) are the configuration parameters of this object. In addition, to allow the users of the library to decide whether to restrict the permissible clients to those specified in the ExternalClientdsTable or not, the BindProcessPolicy configuration parameter is introduced. If the BindProcessPolicy is set equal to ContinueInitialConnection the packets are processed by the unmanageable SmppConnection object which was cloned from the DefaultConnectionTemplate. However, if the BindProcessPolicy is set to SwitchToKnownConnections then after receiving the bind packet, the temporary SmppConnection is deleted and the network connection is assigned to the appropriate manageable SmppConnection within the ExternalClientdsTable table. Figure 4 shows the XML Schema of the SmppServer.
2- SmppConnection: The SmppConnection is a managed class which represents a network connection. According to the SMPP protocol, each connection has the following configuration variables: EnquireLinkTimeout, InactivityTimeout, and ResponseTimeout. We also designate the actions which can be performed on a managed object by an element with minOccurs="0" and maxOccurs="0" in the XML Schema (a technique previously used by [8]) . For the SmppConnection, we have provisioned the Disconnect action. Figure 5 shows the XML Schema of the SmppConnection managed class.
3- SmppExternalClient: The SmppExternalClient represents a valid client (or ESME in SMPP terminology). The SMPP protocol has provisioned the SystemId, SystemType, Password, and PermittedBindTypes configuration parameters for each valid client. It must be mentioned that by overriding the IsPassword method of the SimpleManagedObject class in the SmppExternalClient class we have requested the subagent to store the Password configuration parameter in encrypted format. In addition, each SmppExternalClient includes a child of type SmppConnection which allows the associated connection to be manageable even when the client is disconnected. This manageable SmppConnection object can be used to reconfigure the configuration parameters of a connection and monitor its state, even when the actual networking connection is closed. The proposed trick prevents the monitoring variables associated with a client to be reset after an unwanted disconnection. The schema of this class is depicted in Figure 6.
4- SmppExternalClientsTable: The SmppExternalClientsTable is a tabular managed object which stores the list of the SmppExternalClient objects. The Schema of this class is depicted in Figure 7.

Note that all general configuration properties of an SMPP server-side application are handled by the SMPP networking library itself, without any intervention from the final application. An in-depth experiment on the management of server-side connections of the WordReplacer application is provided as a video file which is accessible at the first author's homepage.

*C. A Sample WordReplacer Application*

In this section, we use the RSCM technology to develop a sample application called WordReplacer. This application contains a SMPP server, which is an instance of the SmppServer class of the previous section, along with a list of words and their replacements. After binding, clients can send messages using SubmitSm packet of the SMPP protocol. When a message contains a word that is in the words list, the word is replaced by its associated replacement word and a new message is composed and returned back to the sender. To show how the RSCM technology allows a modular design, we use the SmppNetworking library of the previous section. From the above definition of the WordReplacer application, we can identify the following classes:
1- WordReplacer which is the main class of the application and represents the whole WordReplacer application. This class has two children: one of type WR_Server (which inherits from the SmppServer) and another child of type WordReplacementRulesList where its definition follows. In addition, it is possible to run the ReplaceWord action on the WordReplacer. The XML schema of the WordReplacer is shown in Figure 8.
2- WR_ExternalClient which inherits from the SmppExternalClient and has an additional configuration parameter called WordReplacementLicensePerMinute. Note how a general SMPP configuration parameter, like SystemId, is specified in the SmppNetworking library while this application-specific configuration parameter is specified in the application-level class WR_ExternalClient (Goals 2 and 3). The XML schema of the WR_ExternalClient class is shown in Figure 9
3- WordReplacementRulesList which contains the list of the replacement rules. Figure 10 shows the XML schema of this class.
4- WordReplacementRule which represents a rule. It contains two configuration parameters: OriginalWord and NewWord. Each rule has a monitoring variable which counts the number of times that this rule has been applied. The XML Schema of this class is depicted in Figure 11.

In addition to the above classes, there are some other classes which inherit from the classes of the SmppNetworking library but have no additional configuration parameters. These are application-specific classes which implement or override some of the functionalities of their parent classes. These classes are:
1- WR_Server which inherits from the SmppServer class.
2- WR_Connection which inherits from the SmppConnection class.
3- WR_ExternalClientsTable which inherits from the SmppExternalClientsTable class.

All the XML Schemas mentioned up to now are associated with some class. However, there should be an XML Schema which designates the format of the XML configuration file as a whole. This XML Schema is depicted in Figure 12. The class diagram of the WordReplacer application along with the participating classes from the SmppNetworking and the RSCM subagent are illustrated in Figure 13. An experiment in which a high-load connection is safely and almost immediately deleted is provided as a video file which is accessible at the first author's homepage.

*D. Managing the WordReplacer Application by a CLI*

We have developed a CLI with a syntax which is compatible with the ITU-T Z.315 standard. This CLI is completely general and can be used to manage any application which is developed by the RSCM technology. In this section we report our experiment in managing the WordReplacer application with this CLI. Figure 14 shows a sample run of the CLI for creating a new managed object. Figure 15 shows another sample run for deleting managed objects from the WordReplacer application (Goals 4 and 5).

*E. Managing WordReplacer Application by a Web-Based Manager*

PeykAsa Company has developed a web-based management system based on the 3GPP 32.603 standard which uses the CORBA technology for the manager-agent communication. We added the RSCM management protocol to the agent of this web-based management system and used the modified system for web-based management of the telecommunication applications developed by the RSCM technology. The user interface of the web-based manager is automatically generated from a database which in turn is filled in automatically using the XSD files of the application.
Since the RSCM technology organizes the managed objects in a tree structure, it seems natural to show a tree view of the managed objects in the user interface. However, the tree representation of a tabular managed object with many children is not readable by human users. In addition, implementing a web interface to add a new tree-structured managed object is a complicated task. Since tabular objects can have only one type of child, a natural representation is to display the subtree of a tabular managed object as a table with each child in a row and each configuration variable in a column. We map the tree structure of the child class to a linear list by adding the name of the parent objects to the beginning of the name of each configuration variable, using dot to join the names. Figure 16 shows the tree representation of a tabular managed object as displayed by our CLI. The tabular representation of the same object in the web-based UI is shown in Figure 17. Note how the configuration variables of the SmppExternalClient class are linearized in the web-based representation. Experiments on performing CREATE and DELETE commands with CLI and web-based managers are recorded in a video file which is accessible at the first author's homepage.

## IV. CONCLUSION

In this paper, we introduced a novel technology called RSCM for runtime reconfiguration of telecommunication applications according to the common requirements of management standards like 3GPP TS 32.602[2] ,and NETCONF [3]. The contributions of the RSCM technology are as follows:

1- Supporting the CREATE and DELETE reconfiguration commands on managed objects within an application.
2- Introducing a novel programming approach which guarantees the safe and almost immediate deletion of managed objects.
3- Using the inheritance feature of the XML technology to extend the object oriented design principle to the configuration file.
4- Supporting the object-oriented development of runtime reconfigurable software libraries, in addition to executable applications.
5- Complete handling of the configuration file. Loading the application according to the configuration file, enforcing the validity of the configuration file, and updating the configuration file according to the latest reconfiguration commands.

The successful use of the RSCM technology in development of commercial telecommunication applications proves the usefulness of the proposed technology.

## V. ACKNOWLEDGMENT

The authors thank their fellows in PeykAsa Company for their help in producing this work. Especially, we want to thank Mr. Hamid Yousefzadeh, the technical manager, for providing the possibility to publish the results of this research, and Saeed Zamani for his help in developing the CLI. The authors also thank Abbas Rasoolzadegan and Samad Paydar for valuable discussions on software engineering issues and are grateful to the anonymous reviewers for their valuable comments which greatly improved the paper.

Table 1: Feature-based comparison between RSCM and related methods.

| Technology/standard implementation  Feature | JMX | Work of Menten | Fractal | libnetconf | RSCM |
|---|---|---|---|---|---|
| Accepting CREATE/DELETE reconfiguration commands from managers. | No | Yes | No | Yes | Yes |
| CREATE/DELETE reconfiguration commands are handled automatically by subagent | No | No | No | No | Yes |
| Mapping the name of a managed object to a reference to the actual entity | Yes | Yes | Yes | No | Yes |
| Automatic handling of the configuration file | No | Yes | Yes | Yes | Yes |
| Distributed specification of validation constraints | No | Yes | Yes | No | Yes |
| Supporting inheritance relation among Managed Objects in the configuration file | No | No | No | No | Yes |
| Appropriate programming interfaces for CREATE/DELETE runtime reconfiguration commands. | No | No | Yes | No | Yes |
| Provisioning the development of manageable software libraries with built-in CREATE/DELETE runtime-reconfigurability. | No | No | Yes | No | Yes |
| Providing a mechanism for [safe and ]almost immediate deletion of managed objects | No | No | No | No | Yes |
| Object Oriented Design | Yes | No | Yes | No | Yes |
| Supporting both C++ and Java programming languages | No | Yes | Yes | No | Yes |

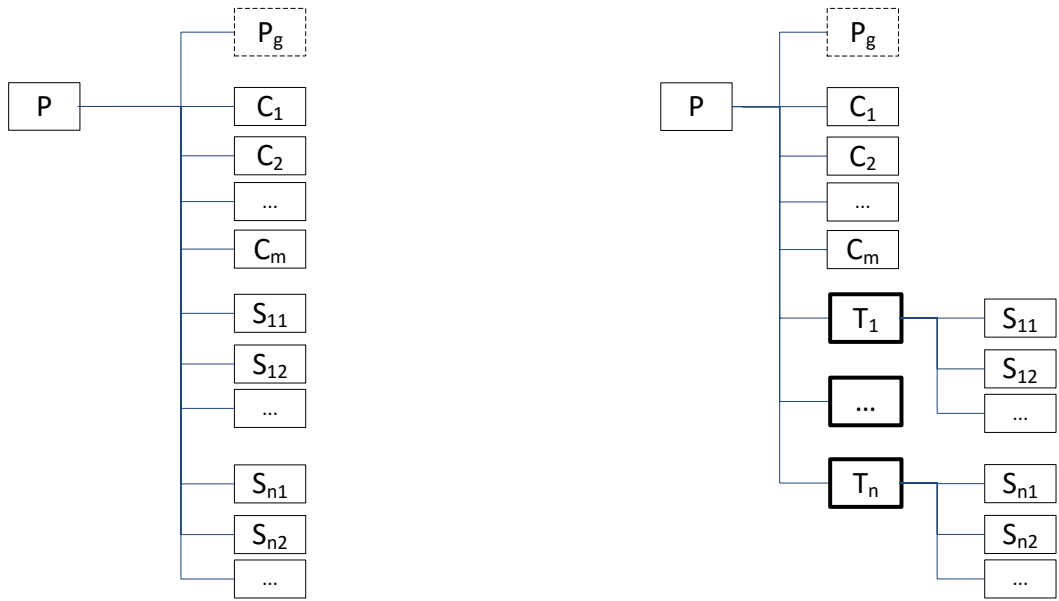

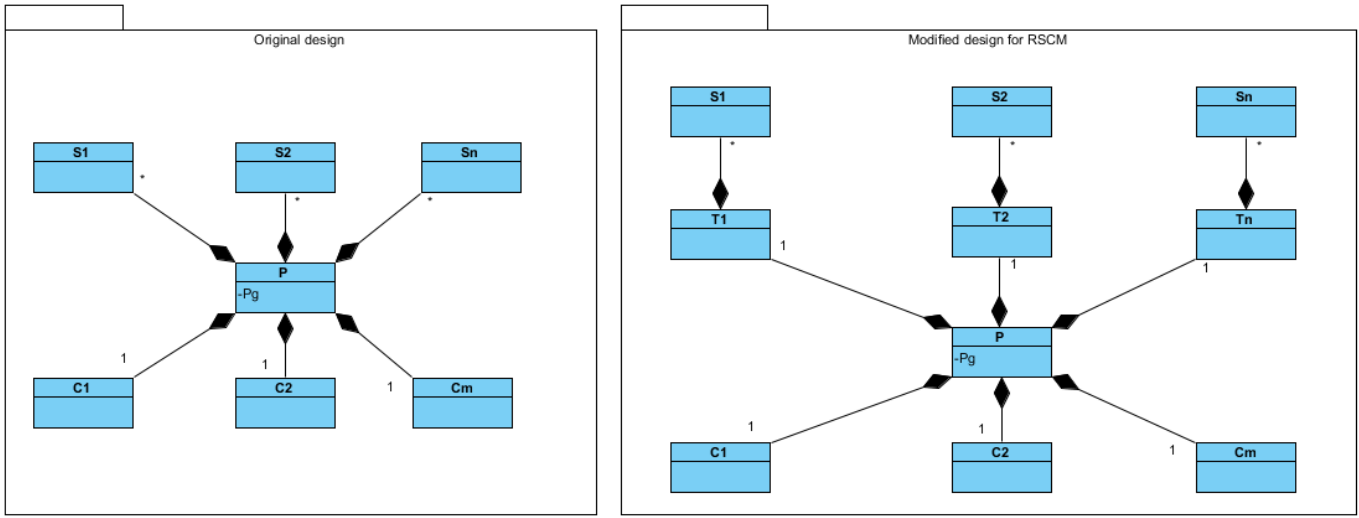

**Figure 1:** Left: Initial unorganized design. The managed class P is composed of a set of configuration parameters $P_g$, a set of fixed children $C_1,\ldots,C_m$, and an arbitrary number of children of types $S_1,\ldots,S_n$. Right: The corresponding design in which managed objects are categorized into Simple and Tabular. The modified class P is a simple managed class with the configuration parameters $P_g$ and fixed children $C_1,\ldots,C_m$ and $T_1,\ldots,T_n$. For $1 \leq j \leq n$, $T_j$ is a tabular managed class containing arbitrary number of children of type $S_j$. In top figures, simple managed objects are depicted by boxes with thin lines, tabular managed objects are depicted by boxes with thick lines, and the configuration variables are depicted by boxed with dashed lines.

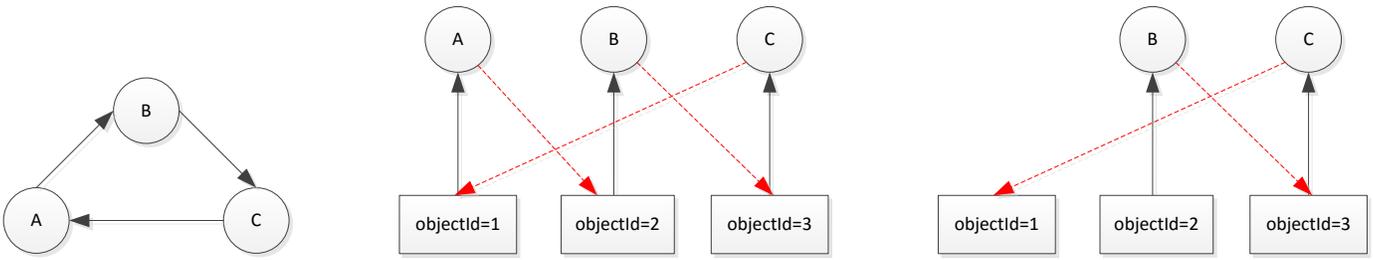

**Figure 2: Left:** Initial direct referencing between objects. **Middle:** Indirect referencing in RSCM. Solid lines indicated ordinary references and dashed lines indicate reference by ObjectId. **Right:** Indirect referencing after deletion of object A. If object C attempts to get a pointer to the deleted object with ObjectId=1, it would receive a null pointer. What happens after object C is notified of deletion of object A depends on the logic of the application.

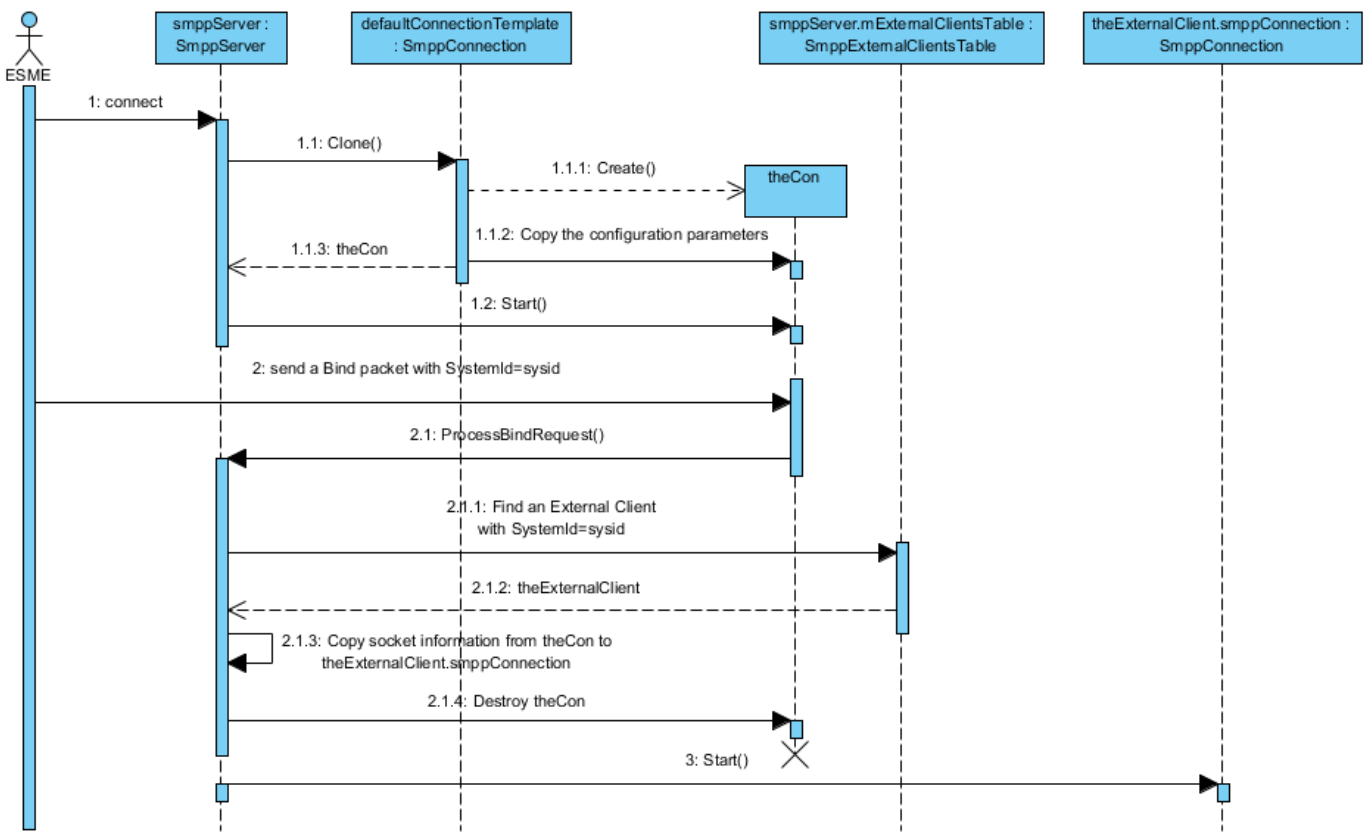

**Figure 3: A Sequence diagram for the scenario in which an ESME connects to an SmppServer.**

```xml
<xs:complexType name="SmppServer">
  <xs:sequence>
    <xs:element name="Port" minOccurs="1">
    <xs:simpleType>
    <xs:restriction base="xs:unsignedShort">
      <xs:minInclusive value="1024" />
    </xs:restriction>
    </xs:simpleType>
    </xs:element>
    <xs:element name="SystemId" type="xs:string" minOccurs="1" />
    <xs:element name="SessionInitTimeout" type="xs:unsignedInt" minOccurs="1" />
    <xs:element name="BindProcessPolicy" minOccurs="1">
      <xs:simpleType>
        <xs:restriction base="xs:string">
          <xs:enumeration value="ContinueInitialConnection" />
          <xs:enumeration value="SwitchToKnownConnections" />
        </xs:restriction>
      </xs:simpleType>
    </xs:element>
    <xs:element name="DefaultConnectionTemplate" type="SmppConnection" minOccurs="1" />
    <xs:element name="ExternalClientsTable" type="SmppExternalClientsTable" minOccurs="1" >
    <xs:key name="ExternalClientSystemIdKey">
        <xs:selector xpath=".//ExternalClient" />
        <xs:field xpath="SystemId" />
    </xs:key>
    </xs:element>
  </xs:sequence>
</xs:complexType>
```

**Figure 4: XML Schema for SmppServer**

```xml
<xs:complexType name="SmppConnection">
    <xs:sequence>
        <xs:element name="EnquireLinkTimeout" type="xs:integer"/>
        <xs:element name="InactivityTimeout" type="xs:integer"/>
        <xs:element name="ResponseTimeout" type="xs:integer"/>
        <xs:element minOccurs="0" maxOccurs="0" name="ActionList">
            <xs:complexType>
                <xs:all>
                    <xs:element name="Disconnect">
                        <xs:complexType>
                            <xs:all>
                            </xs:all>
                        </xs:complexType>
                    </xs:element>
                </xs:all>
            </xs:complexType>
        </xs:element>
    </xs:sequence>
</xs:complexType>
```

**Figure 5: XML Schema for SmppConnection**

```xml
<xs:complexType name="SmppExternalClient">
    <xs:sequence>
      <xs:element name="SystemId" type="xs:string" />
      <xs:element name="SystemType" type="xs:string" />
      <xs:element name="Password" type="xs:string" />
<xs:element name="PermittedBindTypes" default="TRX">
  <xs:simpleType>
  <xs:restriction base="xs:string">
        <xs:enumeration value="TX" />
  <xs:enumeration value="RX" />
  <xs:enumeration value="TRX" />
        </xs:restriction>
</xs:simpleType>
</xs:element>
<xs:element name="Connection" type="SmppConnection" />
    </xs:sequence>
</xs:complexType>
```

**Figure 6: The XML Schema of the SmppExternalClient class**

```xml
<xs:complexType name="SmppExternalClientsTable">
  <xs:sequence>
    <xs:element name="ExternalClient" type="SmppExternalClient" minOccurs="0" maxOccurs="unbounded">
    </xs:element>
  </xs:sequence>
  <xs:attribute name="key" type="xs:string" use="required" />
</xs:complexType>
```

**Figure 7: The XML Schema of the SmppExternalClientsTable class**

```xml
<xs:complexType name="WordReplacer">
  <xs:sequence>
    <xs:element name="WordReplacerServer" type="SmppServer" />
    <xs:element name="WordReplacementRulesList" type="WordReplacementRulesList">
      <xs:key name="OriginalWord">
        <xs:selector xpath=".//WordReplacementRule" />
        <xs:field xpath="OriginalWord" />
      </xs:key>
    </xs:element>
    <xs:element minOccurs="0" maxOccurs="0" name="ActionList">
      <xs:complexType>
        <xs:all>
          <xs:element name="ReplaceWord">
            <xs:complexType>
              <xs:all>
                <xs:element name="Word" type="xs:string" />
              </xs:all>
            </xs:complexType>
          </xs:element>
        </xs:all>
      </xs:complexType>
    </xs:element>
  </xs:sequence>
</xs:complexType>
```

Figure 8: The XML Schema for WordReplacer class.

```xml
<xs:complexType name="WR_ExternalClient">
  <xs:complexContent>
    <xs:extension base="SmppExternalClient">
      <xs:sequence>
        <xs:element minOccurs="1" name="WordReplacementLicensePerMinute" type="xs:unsignedInt" />
      </xs:sequence>
    </xs:extension>
  </xs:complexContent>
</xs:complexType>
```

Figure 9: The XML Schema for WR_ExternalClient

```xml
<xs:complexType name="WordReplacementRulesList">
    <xs:sequence>
   <xs:element name="WordReplacementRule" type="WordReplacementRule" minOccurs="0" maxOccurs="unbounded"/>
    </xs:sequence>
    <xs:attribute name="key" type="xs:string" use="required" />
</xs:complexType>
```

Figure 10: The XML Schema for WordReplacementRulesList

```xml
<xs:complexType name="WordReplacementRule">
    <xs:sequence>
        <xs:element name="OriginalWord" type="xs:string" />
        <xs:element name="NewWord" type="xs:string" />
    </xs:sequence>
</xs:complexType>
```

Figure 11: The XML Schema for WordReplacementRule class.

```xml
<?xml version="1.0" encoding="utf-8" standalone="no" ?>
<xs:schema xmlns:xs="http://www.w3.org/2001/XMLSchema"
attributeFormDefault="unqualified" elementFormDefault="qualified"
xmlns:xsi="http://www.w3.org/2001/XMLSchema-instance">

  <xs:element name="PA_Component">
    <xs:annotation>
      <xs:appinfo>
        <managedElementType value="WordReplacer"/>
      </xs:appinfo>
    </xs:annotation>
    <xs:complexType>
      <xs:all>
        <xs:element name="WordReplacer" type = "WordReplacer"/>
      </xs:all>
    </xs:complexType>
  </xs:element>
</xs:schema>
```

Figure 12: The XML Schema of the whole WordReplacer program.

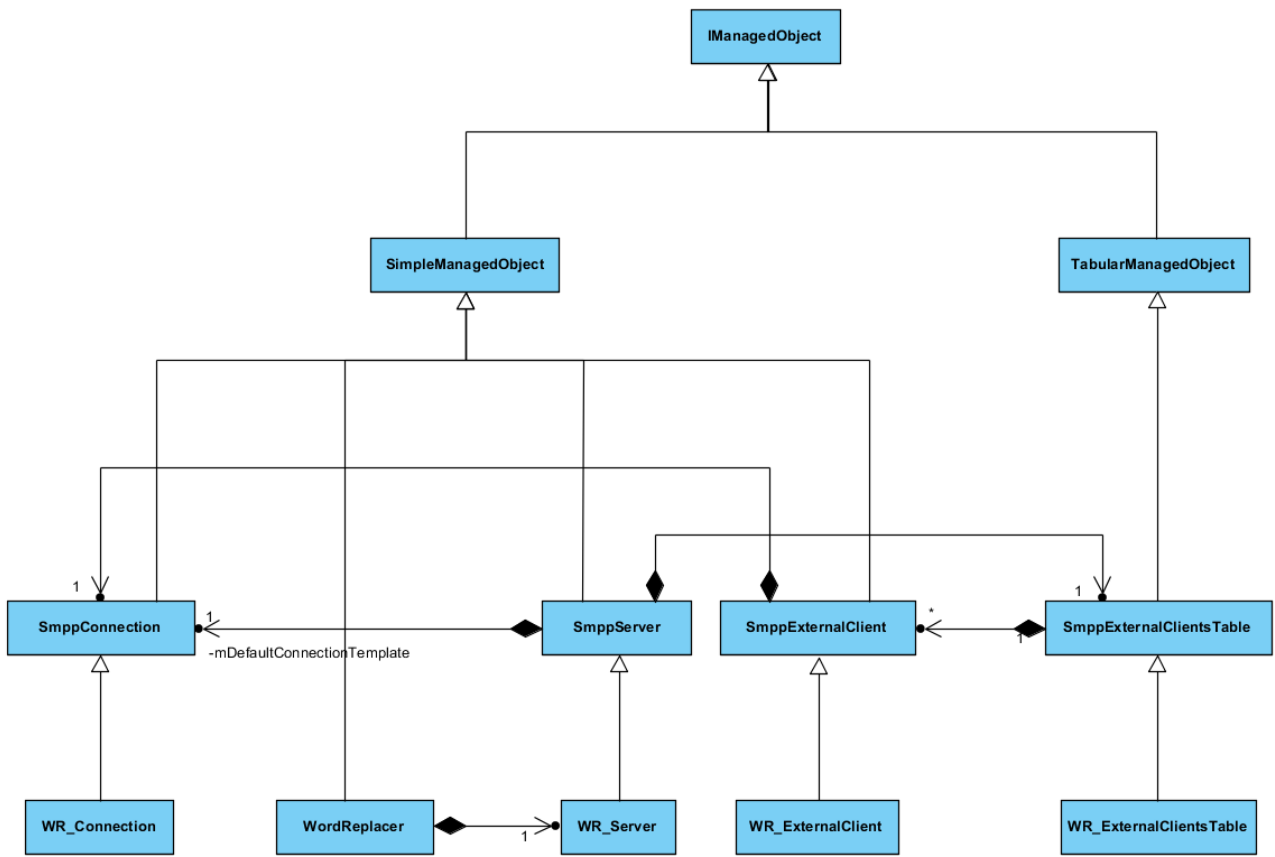

Figure 13: A Class diagram representing some important classes participating in the WordReplacer application. Classes at the two top levels belong to the RSCM library. At the next layer comes the classes belonging to the SmppNetwroking library. The last layer depicts classes that are specific to the WordReplacer application.

```
WR>$help: sig;
        "sig" command prints the signature for adding a new managed object to a parent tabular node.
        Syntax:"sig: <RDN>;" where <RDN> is the relative distinguished name of the parent node (which is a tabular node).
WR>$help: add;
        "add"  command is used to add a child node to another node. The parent node should be a tabular Managed Object.
        Syntax:"add: <RDN>: <MOC>: <GATR>=<VAL>(,<GATR>=<VAL>)*;" where:
        <RDN>  is the relative distinguished name of the parent node
        <MOC>  is the managed object class of the new node
        <GATR> is a generalized attribute of the new node
        <VAL>  is a numeric or string value
        Definition: A generalized attribute of a node is either an attribute of that node or an attribute of a desendent node.
                In the latter case, the name of the generalized attribute consists of the sequence of intermediate nodes
                and the attribute name separated by comma.
        Example: Assume that object T1 is tabular and can contain objects of class C1.
                Objects of class C1 contain and attribute c1 and two objects named A and B.
                In addition, A contains two attributes a1 and a2 and B contains attributes b1 and b2. Then the command
                        "add: T1: C1: c1=7, A.a1=a1, A.a2=2, B.b1=8, B.b2=9;"
                adds a new node of type C, with the specified values for attributes of itself and its children, to T1.
WR>$cn: WordReplacer, WordReplacementRulesList;
WR>WordReplacer,WordReplacementRulesList>$ls;
        WordReplacementRule="acheive"
        WordReplacementRule="accross"
        WordReplacementRule="appearence"
        WordReplacementRule="begining"
        WordReplacementRule="beleive"
WR>WordReplacer,WordReplacementRulesList>$sig;
        MOC (Managed Object Class) of children is: WordReplacementRule.
        The following table lists the attributes of this class.

        Name                                                                    | Type       |Opt/Req   |Default Value
        ------------------------------------------------------------------------+------------+----------+---------------
        OriginalWord                                                            |string      |   R      |
        NewWord                                                                 |string      |   R      |
WR>WordReplacer,WordReplacementRulesList>$add: : WordReplacementRule: OriginalWord="concious", NewWord="conscious";
        Add command succeeded.
WR>WordReplacer,WordReplacementRulesList>$ls;
        WordReplacementRule="acheive"
        WordReplacementRule="accross"
        WordReplacementRule="appearence"
        WordReplacementRule="begining"
        WordReplacementRule="beleive"
        WordReplacementRule="concious"
WR>WordReplacer,WordReplacementRulesList>$lsatr: WordReplacementRule="concious";
        OriginalWord : concious
        NewWord : conscious
```

**Figure 14: A sample run of the CLI used for adding managed objects to an application. First the help command explains the 'sig' and 'add' commands. Then the 'cn' (change node) command is used to select the 'WordReplacer,WordReplacementRulesList' as the active managed object. Then the 'ls' command is used for listing the children of the active managed object. The 'sig' command informs us that the active managed object is a tabular managed object which supports addition of children of type 'WordReplacementRule' and that the 'WordReplacementRule' managed class has two required string attributes named OriginalWord and NewWord. Then a new 'WordReplacementRule' is added to the active managed object. The next 'ls' command shows that the created managed object has been successfully added to the active managed object. Finally, the 'lsatr' (list attributes) command shows that the attributes of the added managed object are set to the true values.**

```
WR>$help: del;
    "del" command is used to delete a child node from a tabular node.
    Syntax: "del: <RDN>;" where <RDN> is the relative distinguished name of the node which is to be deleted. Note that <RDN> cannot be empty
    Example:
        del:M1,M2,M3="id";
WR>$cn: WordReplacer , WordReplacementRulesList;
WR>WordReplacer,WordReplacementRulesList>$ls;
    WordReplacementRule="acheive"
    WordReplacementRule="accross"
    WordReplacementRule="appearence"
    WordReplacementRule="begining"
    WordReplacementRule="beleive"
    WordReplacementRule="buisness"
WR>WordReplacer,WordReplacementRulesList>$del: WordReplacementRule="accross";
    Delete command succeeded.
WR>WordReplacer,WordReplacementRulesList>$ls;
    WordReplacementRule="acheive"
    WordReplacementRule="appearence"
    WordReplacementRule="begining"
    WordReplacementRule="beleive"
    WordReplacementRule="buisness"
WR>WordReplacer,WordReplacementRulesList>$ls: .., WordReplacerServer, ExternalClientsTable;
    ExternalClient="WRClient1"
    ExternalClient="WRClient2"
WR>WordReplacer,WordReplacementRulesList>$del: .., WordReplacerServer, ExternalClientsTable, ExternalClient="WRClient2";
    Delete command succeeded.
WR>WordReplacer,WordReplacementRulesList>$ls: .., WordReplacerServer, ExternalClientsTable;
    ExternalClient="WRClient1"
WR>WordReplacer,WordReplacementRulesList>$
```

**Figure 15:** A sample run of the text-based manager used for deleting managed objects from the WordReplacer application. First the help command explains the 'del' command. Then the 'cn' (change node) command is used to select the 'WordReplacer,WordReplacementRulesList' as the active managed object. Then, the WordReplacementRule with OriginalWord="accross" is deleted. Then, the 'ls' command is executed to verify that the intended object is actually deleted. The subsequent commands delete an ExternalClient from the application.

```
.
--ExternalClientsTable
    |--ExternalClient="WRClient1"
    |   |   SystemId=WRClient1
    |   |   SystemType=Telecom
    |   |   Password=*****
    |   |   PermittedBindTypes=TRX
    |   |   WordReplacementLicensePerMinute=5
    |   |--Connection
    |       EnquireLinkTimeout=60
    |       InactivityTimeout=600
    |       ResponseTimeout=30
    |--ExternalClient="WRClient2"
        |   SystemId=WRClient2
        |   SystemType=Telecom
        |   Password=*****
        |   PermittedBindTypes=TRX
        |   WordReplacementLicensePerMinute=500
        |--Connection
            EnquireLinkTimeout=60
            InactivityTimeout=600
            ResponseTimeout=30
```

**Figure 16:** The tree representation of a tabular managed object used by the CLI.

| SystemId | SystemType | Password | PermittedBindTypes | WordReplacementLicensePerMinute | Connection.EnquireLinkTimeout | Connection.InactivityTimeout | Connection.ResponseTimeout |
|---|---|---|---|---|---|---|---|
| WRClient1 | Telecom | ***** | TRX | 5 | 60 | 600 | 30 |
| WRClient2 | Telecom | ***** | TRX | 500 | 60 | 600 | 30 |

of 1   1-2 of 2

Add   Delete   Update   Select Node   Close

**Figure 17:** The tabular representation of the tabular managed object of Figure 16 in the web-based manager.